
\input phyzzx.tex
\def\solid{(---------)}
\def\dashes{($-~-~-~-$)}


\def\gev{~{\rm GeV}}

\def\fbi{~{\rm fb}^{-1}}


\def\prdj#1{{\it Phys. Rev.} {\bf D{#1}}}
\def\npbj#1{{\it Nucl. Phys.} {\bf B{#1}}}
\def\prlj#1{{\it Phys. Rev. Lett.} {\bf {#1}}}
\def\plbj#1{{\it Phys. Lett.} {\bf B{#1}}}
\def\zpcj#1{{\it Z. Phys.} {\bf C{#1}}}

\def\hl{h^0}
\def\hh{H^0}
\def\ha{A^0}

\def\mhh{m_{\hh}}
\def\mha{m_{\ha}}
\def\hpm{H^{\pm}}
\def\mhpm{m_{\hpm}}
\def\dphi{\delta\phi}
\def\mt{m_t}
\def\mw{m_W}

\def\mz{m_Z}
\def\sinb{\sin\beta}
\def\tanb{\tan\beta}
\def\cosb{\cos\beta}
\def\rta{\rightarrow}
\def\mupmum{\mu^+\mu^-}
\def\taup{\tau^+}
\def\taum{\tau^-}
\def\mtau{m_\tau}
\def\nsd{N_{SD}}
\def\nsdi{\nsd^1}
\def\nsdii{\nsd^2}

\def\anti{\overline}
\def\hn{H}

\def\mhn{m_\hn}

\def\asym{{\cal A}}

\def\epem{e^+e^-}
\def\wm{W^-}
\def\wp{W^+}

\Pubnum{UCD-95-5\cr IFT-3/95\cr}
\date{January, 1995}
\titlepage
\baselineskip 0pt
\titlestyle{{\bf Using Decay Angle Correlations to Detect
CP Violation in the Neutral Higgs Sector}}
\vskip .1in
\centerline{B. Grz\c{a}dkowski}
\vskip .075in
\centerline {\it Institute of Theoretical Physics}
\centerline{\it Department of Physics, Warsaw University,
Warsaw, PL-00-681 Poland}
\vskip .1in
\centerline{J.F. Gunion}
\vskip .075in
\centerline {\it Davis Institute for High Energy Physics}
  \centerline{\it Department of Physics,
University of California, Davis, CA 95616 U.S.A.}

\vskip .3in
\centerline{\bf Abstract}
\baselineskip 0pt

We demonstrate that decay angle correlations in $\taum\taup$
and $t\anti t$ decay modes could allow a determination of whether or
not a neutral Higgs boson is a CP eigenstate. Sensitivity of
the correlations is illustrated in the case of the
$\epem\rta Z \hn$ and $\mupmum\rta \hn$ production processes
for a two-doublet Higgs model with CP-violating neutral sector.
A very useful technique for minimizing `depolarization-factor'
suppressions of the correlations in the $t\anti t$ mode is introduced.
\vskip .4in

Determination of the CP nature of any neutral Higgs boson that is
directly observed will be crucial to fully unravelling the nature of the Higgs
sector.  The Standard Model (SM) Higgs boson is CP-even,
while the Minimal Supersymmetric Model (MSSM) predicts two pure CP-even
and one pure CP-odd neutral Higgs boson. More generally, it is entirely
possible to have either explicit or spontaneous CP violation in the neutral
Higgs sector. Indeed, the simplest non-supersymmetric two-Higgs-doublet
model (2HDM) and the supersymmetric Higgs two-doublet plus singlet model
both allow for Higgs mass eigenstates of impure CP nature.
\Ref\hhg{For a review of Higgs bosons, see
J.F. Gunion, H.E. Haber, G. Kane, S. Dawson, {\sl
The Higgs Hunters Guide}, Addison Wesley (1990).} Here we shall focus
on the 2HDM, in which CP violation results in
three neutral states, $\hn_{i=1,2,3}$, of mixed CP character.

Various types of experimental observables can be considered for
determining the CP character of a given Higgs boson.
The most direct probe is provided by
\REF\bgbslaser{B. Grz\c{a}dkowski and J.F. Gunion, \plbj{294} (1992) 361.}
comparing the Higgs boson production rate
in collisions of two back-scattered-laser-beam
photons of various different polarizations.\refmark{\bgbslaser}\
A certain difference in rates for different photon helicity choices
is non-zero only if CP violation is present, and has a
good chance of being of measurable size for many 2HDM parameter choices.
In the case of a CP-conserving Higgs sector, the dependence
of the $\gamma\gamma\rta \hn$ cross section on the relative orientation
of the transverse polarizations of the two colliding photons
may well allow a determination of whether a given $\hn$ is CP-even or CP-odd.
\REF\gk{J.F. Gunion and J. Kelley, \plbj{333} (1994) 110.}
\REF\kksz{M. Kramer, J. Kuhn, M. Stong, and P. Zerwas, \zpcj{64} (1994) 21.}
\refmark{\gk,\kksz}\ Note that the general utility of the photon-photon
collision polarization asymmetries derives from the fact
that the (one-loop) $\gamma\gamma$ couplings of a CP-odd and CP-even $\hn$ are
similar in magnitude, so that sensitivity is not strongly dependent
upon the CP nature of the $\hn$.

Correlations between decay products can also probe the
CP nature of a Higgs boson. In this paper we focus
on effects that arise entirely at tree-level, \ie\ that
do not rely on imaginary parts generated at one-loop. For the dominant
two-body decays of a Higgs boson, we can define
appropriate observables if we are able to
determine the rest frame of the Higgs boson and if the secondary
decays of the primary final state particles allow
an analysis of their spin or helicity directions. An obvious example
is to employ correlations between the decay planes of the
decay products of $WW$ or $ZZ$ vector boson pairs and/or energy
correlations among the decay products.
\REF\nelson{C.A. Nelson, \prdj{30} (1984) 1937 (E: {\bf D32} (1985) 1848);
J.R. Dell'Aquila and C.A. Nelson, \prdj{33} (1986) 80,93;
\prdj{37} (1988) 1220; \npbj{320} (1989) 86.}
\REF\duncan{M.J.~Duncan, G.L.~Kane and W.W.~Repko, \prlj{55} (1985) 579;
\npbj{272} (1986) 517.}
\REF\ck{D. Chang and W.-Y. Keung, \plbj{305} (1993) 261.}
\REF\ckp{D. Chang, W.-Y. Keung, and I. Phillips, \prdj{48} (1993) 3225.}
\REF\soni{A. Soni and R.M. Xu, \prdj{48} (1993) 5259.}
\REF\bargeretal{V. Barger, K. Cheung, A. Djouadi, B.A. Kniehl and P.M. Zerwas,
\prdj{49} (1994) 79.}
\REF\skjold{A. Skjold, \plbj{311} (1993) 261; \plbj{329} (1994) 305.}
\refmark{\nelson-\skjold}
However, these will not be useful for a purely CP-odd $\hn$ (which
has zero tree-level $WW,ZZ$ coupling and thus decays primarily to $F\anti F$)
or for a mixed-CP $\hn$ in the (most probable) case where the CP-even component
accounts for essentially all of the $WW,ZZ$ coupling strength (thereby yielding
`apparently CP-even' correlations).
In contrast, $\hn$ decays to $\taum\taup$
or $t\anti t$, followed by $\tau$ or $t$ decays, do, in principle,
allow equal sensitivity to the CP-even and CP-odd components of
a given Higgs boson.

Indeed, a $\hn$ eigenstate couples to $F\anti F$ according to:
$ {\cal L}\propto \anti F (a+ib\gamma_5) F \hn\,,$
which yields
$$\langle F_+ \anti F_+ \vert \hn \rangle\propto b+ia\beta_F;\quad
  \langle F_- \anti F_- \vert \hn \rangle\propto b-ia\beta_F\,,\eqn\amps$$
where $\beta_F=\sqrt{1-4m_F^2/\mhn^2}$, helicity-flip amplitudes
being zero. The crucial point is that, in general,
$a$ and $b$ are of comparable magnitude
\REF\knowlesetal{C.D. Froggatt, R.G. Moorhouse, and I.G. Knowles,
\npbj{386} (1992) 63.}
in a CP-violating 2HDM. In the notation of Ref.~[\knowlesetal]
we have
$$\eqalign{a_{t\anti t}=&-{\mt s_1c_3\over v \sinb},\cr
  a_{\taum\taup}=&-{\mtau c_1\over v\cosb},\cr}\quad
  \eqalign{b_{t\anti t}=&-{\mt s_1s_3\cosb\over v\sinb},\cr
  b_{\taum\taup}=&-{\mtau s_1s_3\sinb\over v \cosb},\cr}\eqn\abforms$$
where $v=246\gev$, $\tanb=v_2/v_1$ is the ratio of the vacuum expectation
values
of the neutral Higgs fields that couple to up and down-type quarks,
and $c_1$, $s_1$, $c_3$, $s_3$ are cosines and sines
of neutral Higgs sector mixing angles $\alpha_1$, $\alpha_3$;
the couplings of Eq.~\abforms\ are those that appear in the Euler
angle parameterization of the first Higgs eigenstate as defined in
Ref.~[\knowlesetal]. In this same notation, the $WW$
coupling of $\hn$ is $c_1\cosb + s_1 c_3\sinb$ times the $WW$
coupling of the SM Higgs boson. Note that we have assumed a Type-II
(as defined in Ref.~[\hhg]) 2HDM model, in which charged-lepton couplings
of the $\hn$ are analogous to down-quark couplings.  It is consistent
to take $\hn$ to be the lightest eigenstate and to allow
the angles $\alpha_{1,3}$ appearing in $c_1,s_1,c_3,s_3$ to take on
arbitrary values.\foot{For fixed $\tanb$ and $v$,
there are 7 independent parameters in the
2HDM model in which the discrete symmetry that guarantees
the absence of flavor-changing neutral currents is only softly-broken.
\refmark\knowlesetal\
These can be taken as $\alpha_i$ ($i=1,2,3$), the masses of
the three neutral Higgs bosons, and the mass of the charged Higgs boson.}
Clearly, $a$ and $b$ are generally of similar
size. SM couplings for the $\hn$ are obtained in the case $\alpha_1=\beta$
and $\alpha_3=0$.

The use of azimuthal angular correlations in $\taum\taup$ and $t\anti t$
decays to determine the CP eigenvalue of a pure CP state was explored
in Ref.~[\kksz]; the additional azimuthal angle dependence that
is present only for a mixed-CP eigenstate has been noted in Ref.~[\ckp]
in a special case of $t\anti t$ decays. Here, we shall
present a unified treatment aimed at realistically evaluating
the possibility of using correlations in $\hn\rta \taum\taup$ and $t\anti t$
final states to determine if
a decaying Higgs boson is a mixed CP eigenstate, thereby directly
probing for the presence of CP violation in the Higgs sector.

An efficient framework for our analysis is that developed
\REF\kw{J. Kuhn and F. Wagner, \npbj{236} (1984) 16.}
in Refs.~[\kw,\kksz]. Consider the charged current decay $F\rta R f$, where $F$
is a heavy fermion, $f$ is a light fermion whose mass can be neglected,
and $R$ can be either a single particle or a multiparticle state with known
quantum numbers and, therefore, calculable coupling to the charged weak
current.
(Examples are $\tau\rta R \nu$, where $R=\pi,\rho,A_1,\ldots$, and $t\rta W b$,
where $W$ decays to a fermion plus anti-fermion.)  For the $R$'s of interest,
the form of the hadronic current $J_\mu$, deriving from the standard
$V-A$ interaction for the
$J_\mu\equiv\langle R\vert V_\mu - A_\mu \vert 0 \rangle$ coupling,
is completely determined in terms of the final particle momenta.
Using the particle symbol to denote also its momentum, and defining
$$\Pi_\mu=4{\rm Re}\,J_\mu f\cdot J^*-2 f_\mu J\cdot J^*,\quad
  \Pi_\mu^5=2\epsilon_{\mu\rho\nu\sigma} {\rm Im}\,J^\rho J^{*\,\nu} f^\sigma,
\eqn\pidefs$$
all useful correlations in $\hn\rta F\anti F$ decay can
be obtained by employing the quantities
$$\omega=F\cdot(\Pi-\Pi^5),\quad R_\mu=m_F^2(\Pi-\Pi^5)_\mu - F_\mu F\cdot
(\Pi-\Pi^5)\,, \eqn\omegardefs$$
and their $\anti F$ analogues.
In the $F$ rest frame, $R_0=0$, $\vec{R}=m_F^2(\vec\Pi-\vec\Pi^5)$, and
$|\vec{R}|=m_F\omega$. In fact,
$\vec S_F=\vec{R}/(m_F\omega)$ acts as an effective spin direction
($|\vec S_F|^2=1$) when in the $F$ rest frame.

Let us give some illustrative examples. For $\taum\rta \pi^-\nu$
decay, $J_\mu\propto \pi_\mu^-$
and $\vec S_{\taum}=\hat{\pi^-}$ is the unit vector pointing
in the direction of the $\pi^-$'s three momentum (using
angles defined in the $\taum$ rest frame). For $\taum\rta \rho^-\nu\rta
\pi^-\pi^0\nu$, $J_\mu\propto (\pi^--\pi^0)_\mu$, yielding
$\Pi_\mu\propto 4(\pi^--\pi^0)_\mu \nu\cdot(\pi^--\pi^0)+2\nu_\mu m_\rho^2$,
and, thence,
$\vec S_F\propto m_\tau(\vec{\pi^-}-\vec{\pi^0})(E_{\pi^-}-E_{\pi^0})
+  \vec{\nu}m_\rho^2/2\,,$
where the pion energies and directions are defined in the $\taum$ rest frame.
For $t\rta W^+ b\rta l^+\nu b$,
$J_\mu\propto \anti u(\nu)\gamma_\mu (1-\gamma_5)v(l^+)$,
and $\Pi_\mu\propto l^+_\mu \nu\cdot b + \nu_\mu l^+\cdot b$,
$\Pi_\mu^5\propto \nu_\mu l^+\cdot b-l^+_\mu \nu\cdot b $, so that
$\Pi_\mu-\Pi_\mu^5\propto l^+_\mu$, implying $\vec{S_t}=\hat l^+$
in the $t$ rest frame.

If the full $(\Pi-\Pi^5)_\mu$ can be determined on an event-by-event
basis, then we can define
the `effective spin' vectors $\vec S_F$ and $\vec S_{\anti F}$
{\it for each event}, and the distribution of the Higgs decay products takes
the very general form
$$\eqalign{
dN
\propto &\Bigl[(b^2+a^2\beta_F^2)(1+\cos\theta\cos\anti\theta)
        +(b^2-a^2\beta_F^2)\sin\theta\sin\anti\theta\cos(\phi-\anti\phi)\cr
        &-2ab\beta_F\sin\theta\sin\anti\theta\sin(\phi-\anti\phi)
        \Bigr]d\cos\theta d\cos\anti\theta d\phi d\anti\phi\,,\cr}
\eqn\dnform$$
where $\theta,\phi$ and $\anti\theta,\anti\phi$ define the angles
of $\vec S_F$ and $\vec S_{\anti F}$ in the $F$ and $\anti F$ rest frames,
respectively,
{\it employing the direction of $F$ in the $\hn$ rest frame as the
coordinate-system-defining $z$ axis}.
\foot{Defining the 4-vector $S_\mu=R_\mu/(\omega m_F)$,
and similarly for $\anti S_\mu$, the underlying covariant form of the matrix
element squared is
$|{\cal M}|^2 \propto (a^2+b^2)(F\cdot\anti F-m_F^2S\cdot\anti S)+
(a^2-b^2)(F\cdot\anti F S\cdot \anti S-\anti F\cdot S F\cdot\anti S -m_F^2)
-2ab\epsilon_{\alpha\beta\rho\sigma}S^\alpha\anti S^\beta F^\rho\anti
F^\sigma$, with the convention $\epsilon_{0123}=+1$.
We note that this result reduces
to that of Ref.~[\ckp] (except for a difference in the sign
of the $\epsilon$ term) in the case of $t\anti t$ where (using
their notation $\anti l\equiv l^+$)
$S_\mu= (m_F\anti l_\mu/F\cdot \anti l-F_\mu /m_F)$.}
(Note that the $F$ coordinate
axes are to be used in both the $F$ {\it and}
$\anti F$ rest frames to define the angles appearing in
Eq.~\dnform.)  To employ Eq.~\dnform\ we
must retain the ability to distinguish $F$ from $\anti F$.
In fact, to determine $\sin(\phi-\anti\phi)$
it is absolutely necessary that we be able to determine
the $F$ and $\anti F$ rest frames, \ie\ their
line of flight in the Higgs rest frame.
We shall return to this issue momentarily.

We note that at one-loop $a$ and $b$ can develop imaginary parts.
In this case, $a^2,b^2,ab$ should be replaced in Eq.~\dnform\
by $|a|^2,|b|^2,{\rm Re} (ab^*)$, respectively.  In addition, new
angular dependences arise in Eq.~\dnform\ of the form
$-2{\rm Im}(ab^*)\beta_F(\cos\theta+\cos\anti\theta)$.
In principle, the imaginary parts of $a$ and $b$ are also sensitive
to CP violation in the Higgs sector (but also to other types
of CP violation --- in the SM, non-zero effects appear at the 2-loop
level). After including branching ratios, the statistical
significance associated with isolating
the above term is not very large if the 2HDM provides the only source of CP
violation.\Ref\bohdan{B. Grz\c{a}dkowski \plbj{338} (1994) 71.}
In any case, these extra
terms will not contribute to the correlations upon which we focus.

If we cannot determine $(\Pi-\Pi^5)_\mu$ for each event, then Eq.~\dnform\
must be modified.  An extreme example is $F\rta R f$ decay where the $R$
decay products are not examined.  In this case
the angles of $R$ in the $F$ rest frame would be employed in Eq.~\dnform,
and `depolarization' factors arise as a result of event averaging.
In deriving Eq.~\dnform, the angular independent term is actually
multiplied by $(m_F\omega_F)(m_F\omega_{\anti F})$ and the
$\cos\theta\cos\anti\theta$, $\sin\theta\sin\anti\theta
\sin(\phi-\anti\phi)$ and
$\sin\theta\sin\anti\theta\cos(\phi-\anti\phi)$
terms by $|\vec R_F||\vec R_{\anti F}|$.
On an event-by-event basis the ratio of these coefficients is unity,
as outlined earlier.  When averaged over events, this is no longer true.
Consequently, when event averaging (denoted by $\langle\ldots\rangle$)
all the angle-dependent terms in Eq.~\dnform\ must be multiplied by
$D_F\equiv\langle |\vec R_F|\rangle/(m_F\langle \omega_F\rangle)$ and/or
its $D_{\anti F}$ analogue, relative to the angle-independent term.
We define $D\equiv D_F D_{\anti F}$.

At first sight, the necessity of event averaging arises
in the case of the $t\anti t$ final state, for which we will find that
we must have one top decay leptonically and the other hadronically
in order to define the $t\anti t$ line of flight
and, thereby, appropriate angles in Eq.~\dnform.
For the hadronically decaying top, the problem is to distinguish
the quark vs. anti-quark jet coming from the $W$ so as to construct
$(\Pi-\Pi^5)_\mu$ (which is proportional to the $\wp$ ($\wm$)
anti-quark (quark)
momentum for $t$ ($\anti t$) decay) for each event.\foot{We assume
that there is no efficient technique for determining the
sign of the charges of the quark jets resulting from the top decay.}
If we simply sum over all $W$ decay
product configurations, then the appropriate depolarization factor
is easily computed by using $J_\mu\propto \epsilon^W_\mu$ and
summing over $W$ polarizations.
One finds $\Pi_\mu^5=0$ and $\Pi_\mu\propto b_\mu+W_\mu(m_t^2-\mw^2)/\mw^2$.
Employing this result yields $D_t=(m_t^2-2\mw^2)/(m_t^2+2\mw^2)\sim 0.4$
for $\mt=174\gev$. In the modified Eq.~\dnform\ the angles for the
one hadronically decaying $t$ (or $\anti t$) would then be
those of the $\wp$  (or $\wm$) in the $t$
(or $\anti t$) rest frame. (For the leptonically decaying $\anti t$ (or
$t$) the angles of the $l^-$ (or $l^+$) are directly measured and
the associated $D_{\anti t}$ (or $D_t$) is unity.)
Similarly, if in $\tau\rta R\nu$ the $R$ is spin-1
and its decay products were simply integrated over, a depolarization
factor of $D_{\tau}=(m_\tau^2-2m_R^2)/(m_\tau^2+2m_R^2)$ would enter.

Fortunately, these severe depolarization factors can be avoided
in both cases. For the bulk of $\tau$ decays $R$ is a resonance
of known quantum numbers decaying to easily distinguished particles,
in which case we can construct $(\Pi-\Pi^5)_\mu$ event-by-event
(see, \eg, the $\rho$ example described earlier)
and depolarization factors do not arise. In the case of a hadronically
decaying $t$ ($\anti t$), a simple helicity argument shows that
the most energetic of the $\wp$ ($\wm$) jets in the $t$ ($\anti t$)
rest frame is more likely to be the anti-quark (quark),
\ie\ the equivalent of the $l^+$ ($l^-$). Employing the angles of this most
energetic jet (while integrating over
the angles of all the other jets, so that the angles of this most
energetic jet define the only direction associated with the decay)
yields (via Monte Carlo calculation) $D\sim 0.78$,
essentially independent of $\mhn$ for $\mhn\lsim 800\gev$.

Let us now specify our procedure for isolating the coefficients of the
$\cos(\phi-\anti\phi)$ and $\sin(\phi-\anti\phi)$ angular correlation
terms. Defining $c\equiv \cos\theta$, $\anti c\equiv\cos\anti\theta$,
$s\equiv \sin\theta$, $\anti s\equiv\sin\anti\theta$,
$c_{\phi}\equiv\cos\delta\phi$, $s_\phi\equiv \sin\delta\phi$
(where $\delta\phi\equiv \anti\phi-\phi$), and $d\Omega\equiv dc\, d\anti c \,
d\delta\phi$, and including a possible depolarization factor, we have
$${1\over N}{dN\over d\Omega}={1\over 8\pi} \left[ 1+ D c\anti c
+ \rho_1 s \anti s s_\phi+\rho_2 s\anti s c_\phi\right]\,,\eqn\phidist$$ where
$$\rho_1\equiv D{2ab\beta_F\over (b^2+a^2\beta_F^2)}\,,\quad
\rho_2\equiv D{(b^2-a^2\beta_F^2)\over (b^2+a^2\beta_F^2)}\,.\eqn\rhodefs$$
For a CP-conserving Higgs sector, either $a=0$ or $b=0$ implying
$\rho_1=0$ and $|\rho_2|=D$. For a CP-mixed eigenstate,
both $a$ and $b$ are non-zero. Thus $\rho_1\neq 0$ provides
an unequivocable signature for CP violation in the Higgs sector, while
the difference $D-|\rho_2|$  also provides a measure
of Higgs sector CP violation. (Indeed, $\rho_1$ and $\rho_2$ are not
independent; $\rho_1^2+\rho_2^2=D^2$.) Values of $\rho_1\sim D$
and $\rho_2\sim 0$ are common in an unconstrained 2HDM.

To isolate $\rho_1$ and $\rho_2$, we define projection functions
$f_{1,2}(\theta,\anti\theta,\delta\phi)$ such that
$\int f_{1,2}d\Omega=0$, $\int f_{1,2} c\anti c d\Omega=0$,
$\int f_{1} s \anti s s_\phi d\Omega=8\pi$,
$\int f_{1} s \anti s c_\phi d\Omega=0$,
$\int f_{2} s\anti s s_\phi d\Omega=0$,
and $\int f_{2} s\anti s c_\phi d\Omega=8\pi$.
Then, $\rho_{1,2}=\int f_{1,2} {1\over N}{dN\over d\Omega} d\Omega$.
The critical question
is with what accuracy can $\rho_{1,2}$ be determined experimentally?
In the absence of background, it is easily shown that the
experimental errors of the determination are given by
$\delta\rho_{1,2}=(y_{1,2}-\rho_{1,2}^2)^{1/2}/\sqrt N$, where
$y_{1,2}=\int f_{1,2}^2 {1\over N}{dN\over d\Omega}d\Omega$,
and $N$ is the total number of events. (For the $f_{1,2}$ choices
we shall make, $y_{1,2}=\int f_{1,2}^2 d\Omega/(8\pi)$.)
If background is present, then  this result is modified
to $\delta\rho_{1,2}=[y_{1,2}-\rho_{1,2}^2+(B/S)(y_{1,2}+\rho_{1,2}^2-
2\rho_{1,2}\rho_{1,2}^B)]^{1/2}/\sqrt S$, where $S$ is the
number of $\hn$ events, $B$ is the number of background events,
$\rho_{1,2}^B$ is that for the background alone,
and $\rho_{1,2}$ refers to the signal only.  This result assumes that
the background is precisely known, either by detector Monte Carlo plus theory
or high precision experimental measurement.
The choices $f_1=(8/\pi)\epsilon(s_\phi)$
and $f_2=(8/\pi)\epsilon(c_\phi)$ [where $\epsilon(a)=+1$ ($-1$) for
$a>0$ ($a<0$)] are equivalent to employing simple
asymmetries, and yield (for $B=0$)
$y_{1,2}=(8/\pi)^2$ and $\rho_{1,2}/\delta\rho_{1,2}=
(\pi\rho_{1,2}/8)\sqrt N/[1-(\pi\rho_{1,2}/8)^2]^{1/2}$.
(Note that $(\pi/8)\rho_{1,2}$ are the magnitudes of the
$(N_+-N_-)/(N_++N_-)$-type asymmetries.)
For a functional form expressed in terms of orthogonal
functions (upon integration over $d\Omega$),
the error is minimized by using projection functions which match the
angular dependence of the term of interest.  Thus,
we employ $f_1= (9/2)s\anti s s_\phi$
and $f_2=(9/2)s\anti s c_\phi$, for which $y_{1,2}=9/2$ and
$\rho_{1,2}/\delta\rho_{1,2}=
\sqrt{2/9}\rho_{1,2}\sqrt N/[1-(2/9)\rho_{1,2}^2]^{1/2}$.
Note that $\sqrt{2/9}\sim0.47>(\pi/8)\sim 0.39$.

We now discuss the Higgs production reactions
and the $\taum\taup$ and $t\anti t$ final state decay
modes for which the angles of Eq.~\dnform\ can be experimentally determined.
Consider first the $\taum\taup$ case.  The $\tau$ decays are of two
basic types: $\tau\rta l \nu\nu$ and $\tau \rta R\nu$, where $R$
is a hadronically decaying resonance of known quantum numbers.  Together
these constitute about 95\% of the $\tau$ decays, with $BR(\tau\rta \Sigma
R\nu)\sim 58.8\%$.\Ref\dgroom{D. Groom, PDG private communication.}
In the presence of two or more neutrinos, we cannot determine the $\tau$
rest frame angles without employing a Higgs production reaction in
which the Higgs rest frame can be determined without reference to its decays.
Even if we know
the Higgs rest frame, if either (or both) $\tau$'s decay leptonically
we will still not be able to determine either $\cos\dphi$ or $\sin\dphi$.
Only by knowing the Higgs rest frame and having both $\tau$'s decay to $R\nu$
are there enough constraints to unambiguously determine
$|\dphi|$.\refmark\kksz\
For such decays it is crucial that the charge of $R$ can be determined from an
examination of its decay products; in fact we assume that the $R$
decay products can be fully identified so that $(\Pi-\Pi^5)_\mu$
can be determined for each event, thereby avoiding any depolarization
factor. Thus, we employ $D=1$ and an effective branching
ratio for useful $\taum\taup$ final states of $(0.588)^2$.
Because of a two-fold ambiguity in the kinematic
solution, determination of the sign of $\dphi$ generally
requires vertex tagging of both the $\tau$'s.\Ref\kuhnvertex{J. Kuhn,
\plbj{313} (1993) 458.}\ Thus, the $(0.588)^2$
should also be multiplied by the square of the efficiency for
vertex tagging a $\tau$ when estimating our
ability to measure $\rho_1$; however, because this efficiency is
strongly detector-dependent we do not include it in our
explicit numerical results for $\rho_1$.

In the case of $t\anti t $ decays, we cannot employ purely hadronic
final states for which we would be unable to distinguish $t$ from $\anti t$.
Even if the Higgs rest frame is known (and the four-momenta of both $b$-jets
are measured),
double leptonic decays lead to a two-fold ambiguity in the determination
of $\dphi$, so that only $\cos\dphi$ could be computed.
\foot{The $\sin\dphi$ dependence involving just the two charged leptons
in the double leptonic $t\anti t$ final state pointed out in Ref.~\ckp\
thus cannot be experimentally isolated.}
(As we see from Eq.~\phidist, this is adequate for $\rho_2$.)
Only in the case where one top decays hadronically, and the
other leptonically, are we simultaneously able to determine the
exact $t\anti t$ decay axis and distinguish $t$ from $\anti t$.
Thus, we employ an effective branching ratio for useful $t\anti t$ final
states of $2\times (2/3)\times (2/9)$ (keeping only $l=e,\mu$).
As noted earlier, employing {\it one} hadronic $t$ (or $\anti t$)
decay and identifying
the most energetic jet from the $\wp$ ($\wm$) with the anti-quark (quark)
leads to a depolarization factor of $D\sim 0.78$.
\foot{If the Higgs rest frame is known,
$\rho_2$ can be obtained for double-leptonic decays with
branching ratio $(2/9)^2$ and $D=1$. Combining this channel
with the semi-leptonic channel would result in roughly a 13\%
increase of the statistical
significance values for a CP violation signal
that we shall present in the case of $\rho_2$.}
Finally, we note that for a known Higgs mass and known top mass,
the kinematical constraints in the single-leptonic
$t\anti t$ final states are sufficient
to determine unambiguously the momentum of the single $\nu$, without
knowing ahead of time the Higgs rest frame.
This implies that the
$t\anti t$ final state correlations could in principle be employed at a hadron
collider, although the extra initial and final state radiation present
in hadronic collisions is very likely to lead to too much confusion
for this to work in practice.

In order to assess our ability to experimentally measure $\rho_1$
and $\rho_2$, we have examined $\hn$ production in the reactions
$\epem\rta Z\hn$  at
a future linear $\epem$ collider and $\mupmum\rta \hn$
at a possible future $\mupmum$ collider.
\Ref\mlc{A first general survey of Higgs physics at a $\mupmum$ collider
is being prepared by V. Barger, M. Berger, J. Gunion and T. Han;
see also, the Physics Working Group Summary of the 2nd Workshop on
{\it Physics Potential and Development of $\mupmum$ Colliders},
Sausalito, CA, Nov. 17-19, 1994.}
(In the $Z\hn$ reaction, we employ both hadronic and $l=e,\mu$
charged-leptonic decay modes for the $Z$,
with net total branching ratio of $\sim 76\%$.)
For both reactions, the Higgs rest frame can be determined without reference to
the $\hn$ decays.  For the $\epem$ collider we have adopted
the optimal energy, $\sqrt s=\mz+\sqrt 2 \mhn$, as a function
of Higgs mass, and assumed an integrated luminosity of $85\fbi$.
Our results for statistical significances will assume that this
$Z\hn$ mode is essentially background free.  However, we have
not incorporated any efficiencies for cuts that might be required
to guarantee this. We have also not included the dilution due to
the $ZZ$ continuum background for $\mhn$ values in the vicinity of $\mz$.
Depending upon the detector resolution, this background can be substantial for
$\mhn$ values between about 75 and 105 GeV.  In this interval,
our results should (at best) be considered an upper bound.
For the $\mupmum$ collider we have computed the Higgs signal and the continuum
$\taum\taup$ and $t\anti t$ backgrounds assuming unpolarized beams and
a machine energy resolution
of 0.1\%, with $\sqrt s$ centered at the (already known)
value of $\mhn$.  We adopt an integrated luminosity of $20\fbi$.
This is consistent with the hoped for integrated luminosity of $100-200\fbi$
of a multi-TeV $\mupmum$ collider when run at the lower energies required for
direct production of the $\hn$ in the mass range considered.

For $\mhn$ values such that the $\epem\rta Z\hn$ production mode
is background free, the statistical
significance of a non-zero result for $\rho_1$
is that given earlier, $\nsdi=|\rho_1|/\delta\rho_1$, where
$\delta\rho_1=(9/2-\rho_1^2)^{1/2}/\sqrt{N}$, and $N$ is the number of
events {\it after including the branching ratios required to achieve the
final state of interest}: $BR_{eff}=BR(\hn\rta F\anti F)\times BR(F\anti F\rta
X)$, where the latter $F\anti F$ branching ratios to useful final $X$
states were specified above. In the case of $\rho_2$ we must actually
determine the statistical significance associated with a measurement
of $D-|\rho_2|$.  This is given by
$\nsdii=[D-|\rho_2|]/\delta\rho_2$, where
$\delta\rho_2=(9/2-\rho_2^2)^{1/2}/\sqrt N$.

In $\mupmum \rta \hn$, the continuum backgrounds must be included.
The CP-conserving background does not contribute to $\rho_1$, and
the statistical significance of a non-zero value for
$\rho_1$ is given by $\nsdi=|\rho_1|/\delta\rho_1$ with
$\delta\rho_1=[9/2-\rho_1^2+(B/S)(9/2+\rho_1^2)]^{1/2}/\sqrt S$,
where $S$ is the total number of events from $\hn$ production, and $B$
is the total number of events from the continuum background,
in the final state of interest.
Although the background does have substantial $\cos\dphi$ dependence, not only
should we have an excellent theory plus detector
Monte Carlo simulation of the background, but also
$\rho_2$ for the background
could be directly measured for $\sqrt s$ values on either
side of the Higgs resonance.
Thus, we neglect errors in the background subtraction.
In this case, we have $\nsdii=[D-|\rho_2|]/\delta\rho_2$
with $\delta\rho_2=[9/2-\rho_2^2+(B/S)
(9/2+\rho_2^2-2\rho_2\rho_2^B)]^{1/2}/\sqrt S$,
where $\rho_2^B$ is that for the background alone.
(The non-superscripted $\rho_{1,2}$ are always those of the signal alone.)
Results presented for $\nsdii$
include $\rho_2^B$, but differ negligibly from those
obtained if $\rho_2^B$ is set to zero in the error expression above.

\FIG\asym{The maximum statistical significances $\nsdi$ and $\nsdii$
for $\hn\rta\taum\taup$ \solid\ and $\hn\rta t\anti t$ \dashes,
in $\epem\rta Z\hn$ ($L=85\fbi$) and $\mupmum\rta \hn$ ($L=20\fbi$)
production, after searching over all $\alpha_1$ and $\alpha_3$ values
at fixed $\mhn$ and $\tanb$.  In each case, curves for the three
$\tanb$ values of 0.5, 2, and 20 are shown. In the $\taum\taup$
($t\anti t$) mode $\nsd$ values increase (decrease) with increasing $\tanb$,
except in the case of $\mupmum\rta \hn\rta t\anti t$, where the lowest
curve is for $\tanb=0.5$, the highest curve is for $\tanb=2$,
and the middle curve is for $\tanb=20$.}

\midinsert
\vbox{\phantom{0}\vskip 4.4in
\phantom{0}
\vskip .5in
\hskip -20pt
\special{ insert scr:asym.ps}
\vskip -1.4in }
{\rightskip=3pc
 \leftskip=3pc
\Tenpoint
\baselineskip=12pt
\noindent Figure~\asym:
The maximum statistical significances $\nsdi$ and $\nsdii$
for $\hn\rta\taum\taup$ \solid\ and $\hn\rta t\anti t$ \dashes,
in $\epem\rta Z\hn$ ($L=85\fbi$) and $\mupmum\rta \hn$ ($L=20\fbi$)
production, after searching over all $\alpha_1$ and $\alpha_3$ values
at fixed $\mhn$ and $\tanb$.  In each case, curves for the three
$\tanb$ values of 0.5, 2, and 20 are shown. In the $\taum\taup$
($t\anti t$) mode $\nsd$ values increase (decrease) with increasing $\tanb$,
except in the case of $\mupmum\rta \hn\rta t\anti t$, where the lowest
curve is for $\tanb=0.5$, the highest curve is for $\tanb=2$,
and the middle curve is for $\tanb=20$.
}
\endinsert

Our results for the maximum $\nsdi$ and $\nsdii$ values are presented
in Fig.~\asym, where we have adopted a top quark mass of 174 GeV.
The maximum values were found by searching
over all values for the Euler angles $\alpha_1$ and $\alpha_3$
appearing in Eq.~\abforms, holding $\tanb$ and $\mhn$ fixed.
In general, $\nsdi$ is only slightly larger than $\nsdii$,
as is easy to understand from the fact that $|\rho_1|$ and
$D-|\rho_2|$ both have maximum values close to $D$.
The production rates and branching ratios both depend upon
the couplings of Eq.~\abforms, as well as couplings to other fermions
and to $WW$ and $ZZ$ pairs. Couplings to up-type fermions are, of course,
analogous to the $t\anti t$ coupling given in Eq.~\abforms\ with $\mt$
replaced by the up-type fermion mass, while couplings to down-type fermions
are analogous to the $\taum\taup$ coupling of Eq.~\abforms\ with $\mtau$
replaced by the down-type fermion mass.  In computing the $\taum\taup$
and $t\anti t$ branching ratios, the full set of possible $\hn$
decays to $f\anti f$, $WW$, and $ZZ$ are included.
As noted earlier, the results of Fig.~\asym\ for $\nsdi$ in the
$\taum\taup$ mode do not incorporate efficiencies for $\tau$ vertex tagging,
required to determine the sign of $\dphi$ (as needed for
computing $\rho_1$), and thus
should be multiplied by the efficiency ({\it not} its square)
for $\tau$ tagging. Fortunately,
it is expected that this efficiency will be relatively high for
appropriate detector designs.

Consider first the results for $\epem\rta Z\hn$ collisions.
{}From Fig.~\asym\ we find that detection of CP violation
through both $\rho_1$ and $\rho_2$
is very likely to be possible for $\mhn<2\mw$ via the $\hn\rta\taum\taup$
decay mode. This is an important result given that
various theoretical prejudices suggest that the lightest
Higgs boson is quite likely to be found in this mass range.
For $\mhn$ between $2\mw$ and $2\mt$,
a statistically significant measurement of CP violation will be difficult.
For $\mhn>2\mt$, detecting CP violation in the $t\anti t$ mode
would require a somewhat larger $L$ (of order 5 times
the assumed luminosity of $L=85\fbi$ for $\tanb$ between 2 and 5).

In $\mupmum\rta\hn$ production, Fig.~\asym\ shows that
the maximum $\nsdi$ and $\nsdii$ values
in the $\taum\taup$ mode can remain large out to large Higgs masses
if $\tanb$ is large, but that for small to moderate
$\tanb$ values the statistical significances are
better in $\epem\rta Z\hn$ collisions when $\mhn<2\mw$.
The reason for this is obvious ---
Higgs production in $\mupmum$ collisions is strongly enhanced at
large $\tanb$.  However, Fig.~\asym\ also indicates that
the $\mupmum\rta\hn\rta\taum\taup$
channel has the advantage of possibly small sensitivity
to the $WW$ decay threshold at $\mhn\sim 2\mw$.
Such insensitivity arises when the Euler angles
$\alpha_1,\alpha_3$ are chosen so as to minimize $WW,ZZ$ couplings
(and hence $\hn\rta WW,ZZ$ branching ratios) without sacrificing
production rate. Thus, for $L=20\fbi$ $\mupmum$ collisions could allow
detection of CP violation all the way out to $2\mt$ for $\tanb\gsim 10$.
The $t\anti t$ final state extends the range of $\mhn$
for which detection of CP violation might be possible only somewhat,
and only if $\tanb$ lies in the moderate range near 2.
A factor 10 higher $\mupmum$ luminosity (\ie\ $L=200\fbi$,
requiring a machine design focusing on center-of-mass energies
below 1 TeV and, possibly, several years of running)
would extend the range of possible detection
in both modes: for $\tanb\gsim 2$, $\mhn$ values up to $2\mt$ could be probed
in
the $\taum\taup$ mode, while
the $t\anti t$ mode might be useful out to quite high masses.

Of course, in obtaining the above results we have implicitly assumed
that the $\hn$ does not have additional decays. If it is not
the lightest Higgs eigenstate,
decays of the $\hn$ to a pair of lighter Higgs bosons or to $Z$
plus a lighter Higgs boson might be kinematically allowed.
If present, they would dilute the statistical significances of Fig.~\asym.
However, if decays involving other Higgs eigenstates are significant, then
there are many other
direct `signals' of CP violation in the Higgs sector that could be present.
For example, simultaneous presence of $\hn_2\rta \hn_1\hn_1$ and either
$\hn_2\rta Z\hn_1$ decays (in our notation, $\hn_2$ is the heavier state)
or $\epem\rta Z\rta \hn_1\hn_2$ production
at a significant level would alone
require the $\hn$'s to be a mixture of CP-even (allowing decays
to a pair of Higgs) and CP-odd (allowing $Z$ plus Higgs decay) states.
As another example, a substantial production rate for
$\hn_{1,2}$ in $\epem\rta Z\hn_{1,2}$ combined with either the existence
of $\hn_2\rta Z\hn_1$ decays or $\epem\rta Z\rta \hn_1\hn_2$ production
would imply that the couplings $ZZ\hn_1$, $ZZ\hn_2$ and $Z\hn_1\hn_2$ are all
non-zero, which requires CP violation in the Higgs sector.
\Ref\pomaral{A. Mendez and A. Pomarol, \plbj{272} (1991) 313.}\
\foot{Strictly speaking, the above statements are only true
with regard to tree-level couplings; a $ZZ\hn$ coupling
is present at one-loop even if the $\hn$ is purely CP-odd.
To completely avoid contamination from C-violating one-loop
diagrams, three or more neutral Higgs bosons must be detected.
For example, to all orders, non-zero values
for all three of the couplings $Z\hn_1\hn_2$, $Z\hn_1\hn_3$ and $Z\hn_2\hn_3$
are only possible if CP violation is present.}

We have not explicitly analyzed
the case of a (non-minimal) supersymmetric model with CP violation
in the Higgs sector. However, several comments are useful.
First, decays to superpartner pairs
(neutralino, chargino, slepton, squark pairs) might be important
and would dilute the observability of $\rho_1$ and $\rho_2$ in the $\taum\taup$
and $t\anti t$ channels.  In this case, one can consider using the
the supersymmetric partner pair events themselves.
Generally, a measurement
of $\rho_{1,2}$ in superparticle pair channels would probe a subtle
mixture of Higgs sector CP violation and CP violation deriving from
complex phases in the soft-supersymmetry-breaking
parameters that enter the chargino, neutralino, \etc\ mass matrices.
However, restrictions from neutron and electron electric dipole moments
suggest that the latter phases are quite small, in which case
an observable non-zero value for $\rho_1$ or $D-|\rho_2|$ would
imply CP-violation in the Higgs sector. Procedurally,
those events in which each member of the superparticle pair decays to
jets and/or charged leptons plus a single lightest neutralino (\ie\ the LSP,
of presumably
known mass) would allow treatment along the same lines as the $\taum\taup$
mode after correcting for the finite mass of the single invisible LSP.
Generally, lifetimes would be too short for vertex tagging, and
only $|\dphi|$ could be determined, so that only measurement of $\rho_2$
would be possible. Of course, if the decays of
a supersymmetric model $\hn$ were spread out over many channels, there
might be inadequate statistics in any one channel.

It is also useful to comment on how well $\rho_1$ and $\rho_2$ can be measured
in the case of a CP-conserving Higgs sector.  Recall that
$\rho_1=0$ and $\rho_2=+D,-D$
for a CP-odd, CP-even Higgs boson.  As an example, consider
$\epem\rta Z \hn$, where $\hn$ is CP-even. (Only a CP-even Higgs boson
will have usable $\epem\rta Z\hn$ production rate.) For simplicity,
we assume that the $\hn$ has SM couplings. (The results will then also apply
to the lightest CP-even MSSM Higgs boson $\hl$ in the case where the other
Higgs bosons have masses $\mha\sim\mhh\sim\mhpm\gsim 2\mz$,
in which limit the $\hl$ is SM-like.)
For $L=85\fbi$, we find that the $\taum\taup$ mode yields
$\delta\rho_1\sim\delta\rho_2$ increasing from $\sim 0.05$ to $\sim 0.13$
as $\mhn$ ranges
from 60 GeV up to 160 GeV. Comparing to 1 ($D=1$ in this case),
we see that a simultaneous measurement of
$\rho_1$ and $\rho_2$ would provide a very strong confirmation that
a SM-like Higgs boson is indeed CP-even. $\rho_1$ would be particularly
valuable for $\mhn\sim \mz$ since the $ZZ$ continuum
background yields non-trivial $\cos\dphi$ dependence.
For $\mhn>2\mw$, even a rough determination of $\rho_{1,2}$
would not be possible in this mode due to the rapid fall in
the $\hn\rta\taum\taup$ branching ratio resulting from the onset
of $WW$ and $ZZ$ decays. In the $t\anti t$ mode,
$\delta \rho_1\sim \delta\rho_2$ ranges between 0.4 and 0.8 for $\mhn$
between $2\mt$ and $750\gev$. Comparing to $D\sim 0.78$
we see that statistics would be inadequate to clearly
distinguish between a SM-like CP-even Higgs boson and one of mixed-CP
nature on the basis of the azimuthal angle correlations.
Of course, in a general 2HDM, the $\hn$ in question might be CP-even but have
reduced $WW,ZZ$ coupling.  (In this case, the reduced
production rate via $\epem\rta Z\hn$ would be apparent, but would not
on its own indicate whether the Higgs was CP-even or of mixed-CP character.)
For reduced $WW,ZZ$ coupling,
the $\hn\rta WW,ZZ$ branching fractions decline much more
rapidly than the $Z\hn$ cross section and
the errors on $\rho_1$ and $\rho_2$ in the $t\anti t$ mode
can be sufficiently small to provide a strong
indication of the CP character of the $\hn$.

In summary, our results show that if the Higgs sector is CP-violating then
there is a substantial possibility of explicitly exposing this CP violation
through azimuthal angle correlations between final state particles in
$\hn\rta \taum\taup$, where
the $\hn$ is produced via $\epem \rta Z\hn$ or $\mupmum\rta \hn$,
with assumed integrated luminosities of $L=85\fbi$ and $L=20\fbi$,
respectively. Of particular importance is the general utility of the
$\taum\taup$ mode in $\epem\rta Z\hn$ collisions for Higgs masses in
the theoretically preferred $\mhn\lsim 2\mw$ region.
For $\mhn>2\mt$, azimuthal correlations
in the $t\anti t$ mode also
provide sensitivity to CP violation.  However, statistically
reliable correlation measurements are predicted to be possible
for a smaller portion of parameter space, which, however, would expand
considerably if higher luminosities were available.
The correlations employed rely on the fact that
CP violation generally leads to non-trivial dependence (not
present if the Higgs sector is CP-conserving) on the sine of an
appropriately defined azimuthal angle ($\dphi$), and to dependence
on the cosine of $\dphi$ that is substantially different
than that predicted when CP is not violated (see Eqs.~\dnform\
and \phidist).
Of the two possible CP-violation-sensitive
correlations, $\rho_1$ (obtained
from the $\sin\dphi$ dependence) provides the best opportunity for
detecting CP violation in the 2HDM Higgs sector, both because of
a somewhat larger statistical significance, and because
the associated non-trivial dependence on $\sin\dphi$ cannot arise from
CP-conserving backgrounds or detector efficiency effects.  However,
$\rho_2$ (deriving from dependence on $\cos\dphi$)
provides nearly as good a probe of CP violation.
Further, there is a tendency for both
CP violation `signals' (namely a large value for $\rho_1$ and a small value
for $\rho_2$) to be
simultaneously substantial as a function of the two-Higgs-doublet
model Higgs mixing angle parameters.  On the experimental side, measurement
of $\rho_1$ in the $\taum\taup$ mode requires high efficiency for $\tau$
vertex tagging, in order to determine the sign of the crucial $\dphi$
azimuthal angle, while prospects for measuring both $\rho_1$ and $\rho_2$
in the $t\anti t$ mode could be improved somewhat if a still more efficient
technique for identifying the quark (anti-quark)
jet in $\wm$ ($\wp$) decay were available, \eg\
by determining the charges of the jets coming
from a hadronically decaying $W$.

\bigskip
\centerline{\bf Acknowledgements}
\medskip

This work was supported, in part, by the Department of Energy,
and by the Davis Institute for High Energy Physics.
JFG would like to thank V. Barger and T. Han for collaboration
regarding general Higgs boson production at a $\mupmum$ collider.
BG would like to thank U.C. Davis for support during the course of this
research; his work was also supported in part by the Committee for Scientific
Research under grant BST-475, Poland.

\refout
\bye